\documentstyle{aa}     
\def\ut#1{\mathop{\vtop{\ialign{##\crcr
     $\hfil\displaystyle{#1}\hfil$\crcr\noalign
     {\kern1pt\nointerlineskip}\hbox{$\hfil\sim\hfil$}\crcr
     \noalign{\kern1pt}}}}}

\def\undersymbol#1#2{\mathop{\vtop{\ialign{##\crcr
     $\hfil\displaystyle{#2}\hfil$\crcr\noalign
     {\kern1pt\nointerlineskip}\hbox{$\hfil#1\hfil$}\crcr
     \noalign{\kern1pt}}}}}
\def\arcsec{^{\prime\prime}}

\def\degr{^0}
\begin{document}
\title{The black hole at the galactic center as a possible retro-lens for the S2 orbiting star}
\author{
    F. De Paolis
\inst{1},
       A. Geralico
\inst{1,2},
 G. Ingrosso
 \inst{1} \and
 A. A. Nucita \inst{1} } \offprints{F. De Paolis} \institute{
Dipartimento di Fisica, Universit\`a di Lecce, and {\it INFN},
Sezione di Lecce, CP 193, I-73100 Lecce, Italy \and International
Center for Relativistic Astrophysics - I.C.R.A., University of
Rome ``La Sapienza'', I-00185 Rome, Italy}
\date{Received ; accepted }
\authorrunning{De Paolis et al.}
\titlerunning{The galactic center black hole as a possible retro-lens for the S2 orbiting star}

\abstract{Holz \& Wheeler (\cite{hw}) have recently proposed that
a Schwarzschild black hole may act as a retro-lens which, if
illuminated by a powerful light source, deflects light ray paths
to large bending angles and a series of luminous arcs (or rings in
the case of aligned objects) centered on the black hole may form.
Obviously, the most convenient geometry to get retro-lensing
images would be that of a very bright star close to a massive
black hole, say the putative $\sim 4\times 10^6$ M$_{\odot}$ black
hole at the galactic center. Recent observations of the galactic
center region in the $K$-band have revealed the presence of a very
bright main sequence star (labelled S2) with mass $\sim 15$
M$_{\odot}$ orbiting at close distance ($130-1900$ AU) from Sgr
A$^*$. The relatively vicinity of S2 to the central massive black
hole may offer a unique laboratory to test the formation of
retro-lensing images. The next generation of space-based
telescopes in the $K$-band (like NGST) may have high enough
limiting magnitude necessary to observe such retro-lensing images.

} \maketitle

\section{Introduction}

It is well known that a dark object moving close to the
source-observer line of sight acts as a gravitational lens,
causing the formation of two or more images of the source. If the
images can not be resolved, the lensing phenomenon results in an
overall source image magnification, to which one refers as
gravitational microlensing event. However, the geometry of the
lensing phenomena may also be different with the observer (source)
in between the source (observer) and lens line of sight.

Recently, Holz \& Wheeler (\cite{hw}) have proposed that a
Schwarzschild black hole may act as a retro-lens (or retro-MACHO)
that deflects the source light ray paths to so large bending
angles that emerging photons may reach the observer (in the middle
between the source and the lens). The shape and the magnitude of
the formed images (arcs, becoming rings in the perfect alignment
case) strongly depends on the mass of the black hole and on the
relative distances between the observer, the lens and the source.
In particular, in the perfect alignment case, after $\pi$
rotations around the black hole, photons if collected form a
series of circular rings. More recently, the generalization of the
retro-lensing effect due to Kerr black holes has been analyzed by
De Paolis et al. (\cite{dgin2003}).

Holz \& Wheeler (\cite{hw}) considered the Sun as the light source
which emits photons towards a stellar black hole at distance $D_L$
and then evaluated the magnitude of the formed retro-lensing
images. In the most suitable case (i.e. perfect alignment between
the source, the observer and the lens) the maximum distance at
which the retro images can be seen by an instrument with limiting
magnitude $\bar{m}$ is
\begin{equation}
D_{\rm L} =0.02 {\rm pc} \times \left[10^{(\bar{m}-30)/2.5}(M_{\rm
BH}/10~ M_{\odot})^2\right]^{1/3}~, \label{wheeler}
\end{equation}
for an assumed image baseline of magnitude $m=30$.

However, since we do not expect massive black holes lying so close
to the Earth, we can say that making a survey for retro-MACHOs
illuminated by the Sun might not be a successful strategy.

A better idea is to look towards well known black holes, for
example at the galactic center which is thought to host a super
massive black hole (Sgr A$^*$) with mass $M_{\rm BH}\simeq
4.07\times 10^6 ~M_{\odot}$ (Ghez et al. \cite{ghez2003}).
Moreover, to have a chance to really detect retro-lensing images,
one should also consider a very bright star close to Sgr A$^*$ as
the light-source.

Indeed, if the Sun is the light source and Sgr A$^*$ is the
considered black hole, the magnitude of the formed absorbed
retro-images turns out to be $m_V\simeq 100$ ($\simeq 47$ if
absorption in the $V$ band is not taken into account).

Nowadays, large ground and space-based telescopes allow to obtain
an unprecedent view of the galactic center, in terms of both
sensitivity and angular resolution, so that individual stars can
be monitored within $1\arcsec$ from the galactic center. In this
respect, Ghez et al. (\cite{ghez2003}) have announced the
observation of a bright star, labelled S2, orbiting in close
proximity to the massive black hole at the center of the Galaxy.

The aim of this letter is to show that the binary system S2-Sgr
A$^*$ may offer a unique opportunity to study the retro-lensing
phenomenon proposed by Holz \& Wheeler (\cite{hw}), taking
advantage of both the large black hole mass and the small
separation of the binary components.

\section{S2-Sgr A$^*$ binary system}

\begin{figure}[htbp]
\vspace{5.3 cm} \includegraphics{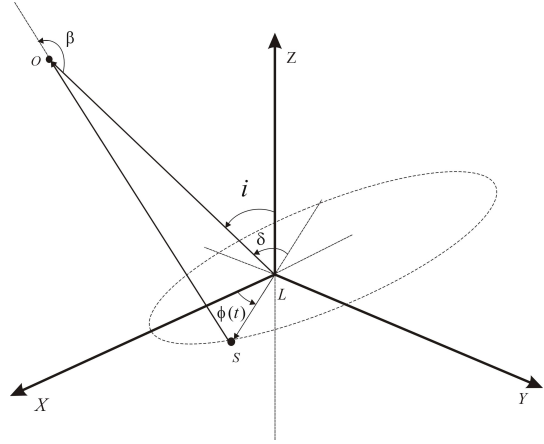} \caption{The geometry of the S2-Sgr A$^*$
binary system retro-lensing phenomenon: photons are emitted by the
source $S$ (the star $S2$), moving on an elliptic orbit (on the
$X-Y$ plane) around the central galactic black hole in $L$. The
emitted photons go towards the lens, which lies in one of two foci
of the ellipse, and after deflection by the angle $\delta$ reach
the observer in $O$.} \label{fig1}
\end{figure}

Very recently, Ghez et al. (\cite{ghez2003}) have observed a star
orbiting in close proximity to the galactic center massive black
hole. The orbital parameters of the binary system are given in
Table \ref{tabella1}. The star, which has been labelled as S2,
with mass $M_{\rm S2}\sim 15$ M$_{\odot}$, appears to be a main
sequence star, probably an O8-B0 dwarf with surface temperature
$\sim 3\times 10^4$ K and bolometric luminosity $\sim 10^3$
L$_{\odot}$.
\begin{table}[h]
\begin{center}
\begin{tabular}{ll}
\hline
\multicolumn{2}{c}{{\rm ~~~S2-Sgr A$^*$ orbital parameters~~~~}}\vspace{0.1 cm}\\
\hline $M_{\rm BH}$&\hspace{1.5 cm}$4.07 \times 10^ 6~M_{\odot}$
\vspace{0.1 cm}\\
$M_{\rm S2}$ & \hspace{1.5 cm}$15~M_{\odot}$ \vspace{0.1 cm}\\
$R_{\rm S2}$ & \hspace{1.5 cm}$5.8~R_{\odot}$ \vspace{0.1 cm}\\
$a$& \hspace{1.5 cm}$4.87 \times 10^{-3}$~pc\vspace{0.1 cm}\\
$e$& \hspace{1.5 cm}$0.87$\vspace{0.1 cm}\\
$P$&\hspace{1.5 cm}$15.78$~yr\vspace{0.1 cm}\\
$i$& \hspace{1.5 cm}$47.3$~deg\\
\hline
\end{tabular}
\end{center}
\caption{The masses $M_{\rm BH}$ and $M_{\rm S2}$ of the galactic
center black hole and S2 orbiting star are given. The remaining
orbital parameters are the S2 star radius $R_{\rm S2}$, the
semi-major axis $a$, the eccentricity $e$, the orbital period $P$
and the inclination angle $i$. Data are taken from Ghez et al.
(2003).} \label{tabella1}
\end{table}
In order to evaluate the magnitude of the retro-image we need to
calculate the magnification factor $\mu$ by considering the
geometry of the S2-Sgr A$^*$ binary system. Referring to Fig.
\ref{fig1}, we consider the S2 source which is moving around the
black hole (at the center of the reference frame) on an elliptic
orbit with semi-major axis $a$ and eccentricity $e$. Let be $i$
the inclination angle of the orbital plane with respect to the
observer (lying the $X-Z$ plane), and define the misalignment
angle $\beta$ as the angle between the source-observer line
$\overrightarrow{SO}$ and the observer-lens line
$\overrightarrow{OL}$, as measured at the observer. Hence, the
misalignment angle $\beta$ is given by
\begin{equation}
\beta=\cos ^{-1}
\left[\frac{{\overrightarrow{SO}}\cdot{\overrightarrow{OL}}}{D_{OS}D_{OL}}\right],
\end{equation}
where $D_{OL}$ and $D_{OS}$ are the observer-lens and the
observer-source distances, with
\begin{equation}
D_{OS}(t)=(D_{LS}^2+D_{OL}^2+2D_{LS}D_{OL}\cos\delta)^{1/2}~.
\label{distOS}
\end{equation}
Here, $D_{OL}\simeq 8$ kpc and $D_{LS}$ is the distance between
the lens and the source, i.e. for the considered elliptic orbit
\begin{equation}
D_{LS}(t)=\frac{a(1-e^2)}{1+e\cos\phi (t)},
\end{equation}
being $\phi=\omega t +\phi_0$ and $\omega =[G(M_{\rm BH}+M_{\rm
S2})/a^3]^{1/2}$. The angle $\delta$ appearing in Eq.
(\ref{distOS}) is the deflection angle, as measured at the lens,
needed for a source photon to reach the observer, i.e.
\begin{equation}
\delta(t) = \cos ^{-1} \left[\frac{\overrightarrow{SL}\cdot
\overrightarrow{LO}}{D_{LS}D_{OL}}\right]= \pi -\cos ^{-1}(\cos
\phi \sin i)~. \label{defl}
\end{equation}
Therefore, the angle $\beta$ turns out to be
\begin{equation}
\beta(t)=\pi - \cos ^{-1}
\left[\frac{D_{OL}+D_{LS}\cos\delta}{D_{OS}}\right]~.
\label{disass}
\end{equation}
In the case of the S2 star, photons retro-lensed by Sgr A$^*$
would form two images, of angular extent $\Delta \Theta \simeq 2
\tan ^{-1} {(R_{\rm S2}/D_{OS}\sin \beta)}$, being $R_{\rm S2}$
the S2 radius. Since lensing conserves the surface brightness, the
total image amplification $\mu$ is given by the ratio between the
fraction of sky covered by each image to that covered by the
source disk as seen by the observer.

Thus, following Holz \& Wheeler (\cite{hw}), one gets for the
brighter (B) and fainter (F) image the amplification factors
\begin{equation}
\mu_{\rm B,F}\simeq [b_o^2(\Theta)-b_i^2(\Theta)]\tan ^{-1} \left(
\frac{R_{\rm S2}}{D_{OS}\sin \beta}\right)
 \frac{D_{OS}^2} {\pi D_{OL}^2R_{\rm S2}^2}~,
\label{amplimisali}
\end{equation}
where, for a given deflection angle $\Theta$, $b_o(\Theta)$ and
$b_i(\Theta)$ are the photon impact parameters corresponding to
the inner and outer radii of the images, respectively. Finally,
the total amplification results to be
\begin{equation}
\mu =\mu_B+\mu_F~. \label{totalamplimisali}
\end{equation}

It has been shown by Chandrasekhar (\cite{chandra}) that, for a
Schwarzschild black hole and in the case of large values of
$\Theta$, the photon impact parameter is
\begin{equation}
b(\Theta)=b_c + b_d e^{-\Theta}~,
\label{bSchw}
\end{equation}
where
\begin{eqnarray}
b_c&=&3\sqrt{3}\frac{G M_{\rm BH}}{c^2}~,\nonumber \\
b_d&=&648 e^{-\pi}\sqrt{3}\frac{G M_{\rm
BH}}{c^2}\left(\frac{\sqrt{3}-1}{\sqrt{3}+1}\right)^2~.
\label{impactparameter}
\end{eqnarray}

Since the angular extent of the S2 disk with respect to Sgr A$^*$
is $\alpha \simeq R_{\rm S2}/D_{LS}(t)$, the bending angles for
the brighter and fainter images are $\Theta_{\rm B}=\delta \mp
\alpha$ and $\Theta_F=2\pi-\delta \mp \alpha$, respectively.

\begin{figure}[htbp]
\vspace{7.7 cm} \includegraphics{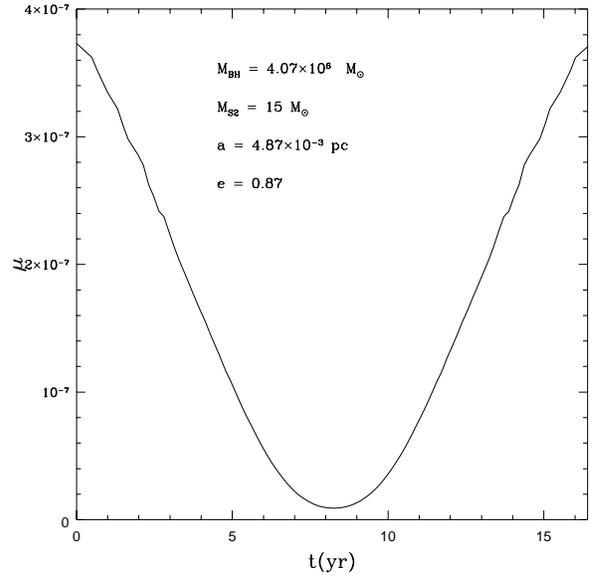} \caption{The amplification factor $\mu$
as a unction of time $t$ is shown. The calculation has been made
by considering the S2-Sgr A$^*$ orbital parameters given in Table
\ref{tabella1}. Note that the periodic behavior of the curve
depends on the S2 orbital motion.} \label{fig2}
\end{figure}

Consequently, by using Eqs. (\ref{amplimisali})-
(\ref{impactparameter}), it is possible to evaluate, for each
position of the star S2 along its orbit around Sgr A$^*$, the
total amplification factor $\mu$ of the retro-lensing images. The
obtained result is shown in Fig. \ref{fig2}. It is evident the
periodic behavior (with period $P$) of the amplification factor
$\mu$, the upper value of which ($\sim 3.75\times 10 ^{-7}$)
corresponds to the periastron distance of S2.

The luminosity of the formed arcs is $L_{\lambda}=\mu
L_{\lambda}^{\rm S2}$, being $L_{\lambda}^{\rm S2}$ the S2
luminosity in the electromagnetic band centered at the wavelength
$\lambda$. It follows that the image magnitude is
\begin{equation}
m_{\lambda}= m_{\lambda}^{\rm S2}-2.5\log \mu~, \label{magnitude}
\end{equation}
being $\mu$ independent on the wavelength.
For S2 in the periastron position ($\phi = 0$ and $D_{LS}\simeq ~
130$ AU) and for the inclination angle $i=47.3 \degr$, from Fig.
\ref{fig2} the maximum allowed amplification factor is $\mu _{\rm
max} \simeq 3.75\times 10^{-7}$ so that the expected minimum image
magnitude results to be $m_{\lambda~{\rm min}}\simeq
m_{\lambda}^{\rm S2}+16.1$.

Then, as a consequence of the S2 orbital motion the amplification
factor $\mu$ decreases until the minimum value $\mu _{\rm min}
\simeq 1.5\times 10^{-8}$ is reached, at the furthest distance
from Sgr A$^*$, corresponding to the maximum image magnitude
$m_{\lambda ~{\rm max}}\simeq m_{\lambda}^{\rm S2}+19.6$.

Clearly, due to the absorption processes along the line of sight,
the previous values have to be increased by the $\lambda$-band
extinction factor $A_{\lambda}$.

The S2 star has been observed in the $K$-band (centered at
$2.2~{\rm \mu m}$) with an unabsorbed magnitude of $m_K^{\rm S2}=
13.9$ (Ghez et al. \cite{ghez2003}). Taking into account the
infrared extinction factor $A_K\simeq 3.3$ towards the galactic
center (Rieke et al. \cite{rieke}), from Eq. (\ref{magnitude}) the
expected absorption corrected $K$-magnitude of the retro-lensing
images \footnote{Since S2 has been classified as a O8-B0 main
sequence star (Ghez et al. \cite{ghez2003}), one expects that the
optical band absolute magnitude is $M_V = -4.0$ (Allen
\cite{allen}) implying, for an assumed observer-S2 distance of $8$
kpc, the apparent $V$-magnitude $m_V^{\rm S2}\simeq 10.6$. Hence,
the expected unabsorbed ring magnitude is in the range
$26.7-30.2$, depending on the S2 orbital position. Unfortunately,
the ratio of the $K$ to $V$ extinction factor is $A_K/A_V\simeq
0.112$ (Rieke \& Lebofsky \cite{rieke1985}) so that the $V$-band
ring magnitude has to be correct for absorption by the factor
$A_V\simeq 29.5$. This implies that the S2 retro-image has
magnitude in the range $56.2\leq m_V\leq 59.7$, i.e. is much
fainter than in the $K$-band.} is
\begin{equation}
m_K=17.2 - 2.5\log \mu~,
\end{equation}
which for the minimum and maximum distances of S2 from Sgr A$^*$
(see Fig. \ref{fig2}), gives the values $m_{\rm K ~ min}\simeq
33.3$ and $m_{\rm K ~ max}\simeq 36.8$, respectively.

Obviously, in order to detect such retro-images, we need
instrument with limiting magnitude $\bar{m} \geq m_{\lambda}^{\rm
S2}+A_{\lambda}-2.5 \log \mu$ in which $A_{\lambda}$ is the
$\lambda$-band extinction factor.

For the sake of completeness, we mention that assuming the
existence of a star (like S2 and with the same orbital parameters)
perfectly aligned with the Earth-Sgr A$^*$ line of sight ($\phi =
0$, $i=\pi/2$ and $\beta =\pi$), the amplification factor and the
corresponding ring magnitude result to be $\mu \simeq 8\times
10^{-4}$ and $m_{\lambda} \simeq m_{\lambda}^{\rm S2}+7.74$,
respectively. In particular, in the $K$-band, after correction for
the absorption, we get a ring magnitude of $m_K\simeq 21.6$, close
to the present day instrument capabilities.

In this case the required instrument limiting magnitude to observe
the retro-lensing images (rings in this case) has to satisfy the
relation
\begin{equation} \bar{m}\geq m_{\lambda}^* +A_{\lambda}-2.5\log
\left[\left(\frac{M}{M_{\odot}}\right)^2\left(\frac{R_{\odot}}{R_S}\right)
\frac{D_{OS}^2}{D_{LS}D_{OL}^2}\right]~.
\end{equation}

\section{Discussion}

Retro-lensing phenomena, i.e. deflections by black holes of the
light ray paths to large bending angles, have been recently
proposed (Holz \& Wheeler \cite{hw}) as a new way to search for
black holes \footnote{A somehow related idea has been proposed by
Capozziello et al. (\cite{capozziello}) who considered the effect
of defocusing gravitational lensing.}. However, if one considers
the Sun as the light source, the limiting distance $D_L$ within
which a Schwarzschild black hole may show observable effects is
given by Eq. (\ref{wheeler}), which assuming $M=10~M_{\odot}$ and
$\bar{m}=30$ gives $D_L\simeq 0.02$ pc. Consequently, due to the
lack of any massive black holes within this distance, making a
survey for retro-MACHOs illuminated by the Sun might not be a
successful strategy.

A better chance to detect retro-lensing images should be looking
towards the direction of well known massive black holes such as
the $\simeq 4.07\times 10^6 ~M_{\odot}$ one (Sgr A$^*$) at the
galactic center.

We have shown that the recently observed bright star S2, shining
the galactic center black hole, may produce retro-lensing images
with apparent magnitude given by Eq. (\ref{magnitude}) in which
$m_{\lambda}^{\rm S2}$ is the S2 star magnitude in the
$\lambda$-band and $\mu$ the amplification factor given by Eq.
(\ref{totalamplimisali}). By taking into account the
$\lambda$-band extinction $A_{\lambda}$, these images may be
observed by an instrument with limiting magnitude $\bar{m} \geq
m_{\lambda}^{\rm S2}+A_{\lambda}-2.5 \log \mu$ which gives, in the
$K$-band, $\bar{m}\simeq 33.3$ at the periastron distance of S2.
Since the photon impact parameter values $b_o$ and $b_i$ are very
close to the Schwarzschild black hole radius value $R_{\rm
Sch}\simeq 1.2 \times 10^{12}$ cm, the retro-lensing images form
at $\simeq 20~\mu$arcsecs from Sgr A$^*$ implying that extremely
high angular resolution are required to see the details of the
retro-images.

As we have seen, the expected retro-lensing image magnitude turns
out to be $\bar{m}\simeq 33$ in the $K$-band for S2 in the
pericenter position. Of course, present instrumentations have not
the right capability to detect such faint images. SIRTF (Space
Infrared Telescope Facility), a space based telescope operating in
the electromagnetic band $3-180$ $\mu$m (National Research Council
\cite{sirtfngst}), will allow to detect, with an integration time
of $3$ hours, a flux $\Phi \simeq 10^{-15}$ erg cm$^2$ s$^{-1}$
corresponding to a limiting magnitude of $\bar{m}\simeq 22$.
Hence, SIRFT has not the required sensitivity to detect the S2
retro-images in the $K$-band. The attainable limiting magnitude
will increase with the Next Generation Space Based Telescope
(NGST) which, planned to work in the wavelength range $0.6-27$
$\mu$m (National Research Council \cite{sirtfngst}), is expected
to have a sensitivity at least 3 order of magnitude (at $2.2$
$\mu$m) better than SIRFT. In fact, the energy flux (in a band
centered at $2.2$ $\mu$m) that NGST is able to detect, within an
integration time of $3$ hours, is $\Phi\simeq 2\times 10^{-19}$
erg cm$^2$ s$^{-1}$ which corresponds to a limiting magnitude
$\bar{m}\sim 32$. This limiting magnitude is very close to that
necessary to observe S2 retro-lensing images that could be
effectively observed by NGST increasing the integration time to
about $27$ hours.

We have also to mention that, in addition to look towards Sgr
A$^*$ in the $K$ band, another possibility to observe
retro-lensing images in the future may be offered by $X$-ray
observations. Indeed, due to $X$-ray interferometry from space,
the next generation of $X$-ray telescope facilities may reach an
angular resolution down to $0.1$ $\mu$arcsec (see e.g. MAXIM
(\cite{maxim}) and Constellation X (\cite{constellationx})) .
Clearly, the problem is the existence of a bright enough $X$-ray
source closely shining the Sgr A$^*$ black hole (S2, being a main
sequence star, does not emit substantially in the $X$-ray band).
Indeed, the galactic center region hosts a large number of $X$-ray
emitting neutron stars or accreting black holes most of which are
detected in the energy range $2-10$ keV with a luminosity
$10^{32}-10^{35}$ erg s$^{-1}$ (Wang et al. \cite{wgl}, Pessah \&
Melia \cite{pm}). Therefore, in the future the detection of
retro-lensing  images in the $X$-ray band may be a challenging
possibility.

A natural question that arises is what changes if the black hole
at the galactic center is a Kerr black hole. Recently, the
generalization of the retro-lensing effect due to rotating Kerr
black holes has been analyzed by De Paolis et al.
(\cite{dgin2003}). Apart a minor effect in changing the value of
the magnitude of the retro-lensing images, the black hole spin
produces a deformation of the image shape since co-rotating and
counter-rotating photons have closer and further impact
parameters, respectively (Peter \cite{peter76}, Bray \cite{bray}).
The analysis of the retro-image shape may therefore allow to
specify the geometry of the system and measure the spin parameter
of the black hole.

\end{document}